\documentclass[12pt,a4paper]{article}
\usepackage{amsmath,amssymb}
\usepackage[dvips]{lscape,graphicx}

\newcommand{\sfrac}[2]{\frac{\scriptstyle #1}{\scriptstyle #2}}

\begin{document}

\author{R. B. Nevzorov${}^{\dag,\ddag}$ and M. A. Trusov${}^{\dag}$ \\[5mm] {\itshape ${}^{\dag}$ITEP, Moscow, Russia} \\
{\itshape ${}^{\ddag}$DESY Theory, Hamburg, Germany}}

\title{Quasi--fixed point scenario in the modified NMSSM}

\maketitle

\begin{abstract}
\noindent The simplest extension of the MSSM that does not
contradict LEP\,II experimental bound on the lightest Higgs boson
mass at $\tan\beta\sim 1$ is the modified Next--to--Minimal
Supersymmetric Standard Model (MNSSM). We investigate the
renormalization of Yukawa couplings and soft SUSY breaking terms
in this model. The possibility of $b$--quark and $\tau$--lepton
Yukawa coupling unification at the Grand Unification scale $M_X$
is studied. The particle spectrum is analysed in the vicinity of
the quasi--fixed point where the solutions of renormalization
group equations are concentrated at the electroweak scale.
\end{abstract}

\newpage

\section{Introduction}

A rapid development of experimental high--energy physics over the
last decades of the XX century gave impetus to intensive
investigations of various extensions of the Standard Model. Its
supersymmetric generalization known as the Minimal Supersymmetric
Standard Model (MSSM) is one of the most popular extensions of the
Standard Model. The Higgs sector of the MSSM includes two doublets
of Higgs fields, $H_1$ and $H_2$. Upon a spontaneous breakdown of
gauge symmetry, each of them develops a vacuum expectation value;
we denote the corresponding vacuum expectation values by $v_1$ and
$v_2$. The sum of the squares of the vacuum expectation values of
the Higgs fields is $v^2=(246\text{~GeV})^2$, the ratio of the
expectation values being determined by the angle $\beta$. By
definition, $\beta=\arctan(v_2/v_1)$. The value of $\tan\beta$ is
not fixed experimentally. It is varied within a wide interval,
from $1.3-1.8$ to $50-60$. Within supersymmetric (SUSY) models,
the upper and lower limits on $\tan\beta$ arise under the
assumption that perturbation theory is applicable up to the scale
at which gauge coupling constants are unified, $M_X=3\cdot
10^{16}\text{~GeV}$ -- that is, under the assumption that there is
no Landau pole in solutions to relevant renormalization group
equations.

The spectrum of the Higgs sector of the MSSM contains four massive
states. Two of them are CP--even, one is CP--odd, and one is
charged. The presence of a light Higgs boson in the CP--even
sector is an important distinguishing feature of SUSY models. Its
mass is constrained from above as
\begin{equation}
m_h\le\sqrt{M_Z^2\cos^2 2\beta+\Delta}\, , \label{1}
\end{equation}
where $M_Z$ is the $Z$--boson mass ($M_Z\approx 91.2\text{~GeV}$)
and $\Delta$ stands for the contribution of loop corrections. The
magnitude of these corrections is proportional to $m_t^4$ ($m_t$
is the running mass of the $t$--quark), depends logarithmically on
the supersymmetry breakdown scale $M_S$, and is virtually
independent of the choice of $\tan\beta$. An upper limit on the
mass of the light CP--even Higgs boson within the MSSM grows with
increasing $\tan\beta$ and, for $\tan\beta\gg 1$, reaches
$125-128\text{~GeV}$ in realistic SUSY models with $M_S\le
1000\text{~GeV}$.

At the same time it is known from \cite{1} that, for $\tan\beta\ll
50-60$, solutions to the renormalization group equations for the
$t$--quark Yukawa coupling constant $h_t(t)$ are concentrated in
the vicinity of the quasi--fixed point
\begin{equation}
Y_{\text{QFP}}(t_0)=\frac{E(t_0)}{6F(t_0)}\, , \label{2}
\end{equation}
where
\begin{gather*}
E(t)=\left[\frac{\tilde{\alpha}_3(t)}{\tilde{\alpha}_3(0)}\right]^{16/9}
\left[\frac{\tilde{\alpha}_2(t)}{\tilde{\alpha}_2(0)}\right]^{-3}
\left[\frac{\tilde{\alpha}_1(t)}{\tilde{\alpha}_1(0)}\right]^{-13/99}
,\quad F(t)=\int\limits_0^t E(t')dt',\\
\tilde\alpha_i(t)=\left(\frac{g_i(t)}{4\pi}\right)^2,\quad
Y_t(t)=\left(\frac{h_t(t)}{4\pi}\right)^2,
\end{gather*}
with $g_i$ being the gauge constants of the Standard Model group.
The variable $t$ is defined in the standard way:
$t=\ln(M_X^2/q^2)$. Its value at the electroweak scale is
$t_0=2\ln(M_X/M_t^{\text{pole}})$, where $M_t^{\text{pole}}\approx
174.3\pm 5.1\text{~GeV}$ is the pole mass of the $t$--quark. Along
with the $t$--quark Yukawa coupling constant, solutions to the
renormalization group equations for the corresponding trilinear
coupling constant $A_t$ characterising the interaction of scalar
fields and the combination $\mathfrak{M}_t^2=m_Q^2+m_U^2+m_2^2$ of
the scalar particle masses also approach the infrared quasi--fixed
point. The properties of solutions to the renormalization group
equations within the MSSM and the spectrum of particles in the
infrared quasi--fixed point regime at $\tan\beta\sim 1$ were
investigated in \cite{2},\cite{3}.

A reduction of the number of independent parameters in the
vicinity of the infrared quasi--fixed point considerably increased
the predictive power of the theory. On the basis of the equation
relating the Yukawa coupling constant for the $t$--quark with its
mass at the electroweak scale
\begin{equation}
m_t(M_t^{\text{pole}})=\frac{h_t(M_t^{\text{pole}})}{\sqrt{2}}v\sin\beta,
\label{3}
\end{equation}
and the value calculated for the running mass of the $t$--quark
within the $\overline{MS}$--scheme ($m_t(M_t^{\text{pole}})=165\pm
5\text{~GeV}$), it was shown in \cite{3}-\cite{5} that, for a
broad class of solutions satisfying the renormalization group
equations within the MSSM and corresponding to the infrared
quasi--fixed point regime, $\tan\beta$ takes values in the
interval between $1.3$ and $1.8$. These comparatively small values
of $\tan\beta$ lead to much more stringent constraints on the mass
of the lightest Higgs boson. A detailed theoretical analysis
performed in \cite{3},\cite{4}, revealed that, in the case being
considered, its mass does not exceed $94\pm 5\text{~GeV}$, which
is $25-30\text{~GeV}$ below the absolute upper limit in the
minimal SUSY model. It should be noted that the LEP\,II
constraints on the mass of the lightest Higgs boson \cite{6} are
such that a considerable fraction of solutions approaching a
quasi--fixed point at $\tan\beta\sim 1$ have already been ruled
out by existing experimental data.

All the aforesaid furnishes a sufficient motivation for studying
the Higgs sector in more complicated SUSY models, as well
renormalization group equations and solutions to these equations
therein. The present article is devoted to an analysis of coupling
constant renormalization within a modified Next--to--Minimal SUSY
Model (MNSSM), where the mass of the lightest Higgs boson can be
as large as $120-130\text{~GeV}$ even at comparatively small
values of $\tan\beta\sim 2$. In addition, the spectrum of
superpartners of observable particles and of Higgs bosons is
studied in the vicinity of the quasi--fixed point of the
renormalization group equations within the MNSSM.

\section{Modified NMSSM}

The Next--to--Minimal Supersymmetric standard Model (NMSSM)
\cite{7}-\cite{9} is the simplest extension of the MSSM.
Historically, the NMSSM arose as one of the possible solutions to
the problem of the $\mu$--term in supergravity (SUGRA) models
\cite{7}. Along with observable superfields, SUGRA theories
contain a hidden sector that includes the dilaton and moduli
fields ($S$ and $T_m$, respectively), which are singlet in gauge
interactions. The total superpotential in SUGRA models is usually
represented as an expansion in the superfields of the observable
sector; that is,
\begin{equation}
W=W_0(S,T_m)+\mu(S,T_m)(\hat{H}_1\hat{H}_2)
+h_t(S,T_m)(\hat{Q}\hat{H}_2)\hat{U}_R^c+\dots, \label{4}
\end{equation}
where $W_0(S,T_m)$ is the superpotential of the hidden sector. The
expansion in (\ref{4}) presumes that the parameter $\mu$ appearing
in front of the bilinear term $(\hat{H}_1\hat{H}_2)$ must be about
the Planck mass, since this scale is the only dimensional
parameter characterising the hidden sector of the theory. In this
case, however, the Higgs bosons $H_1$ and $H_2$ acquire an
enormous mass ($m^2_{H_1,H_2}\simeq\mu^2\simeq M_{\text{Pl}}^2$)
and $SU(2)\otimes U(1)$ symmetry remains unbroken.

In the NMSSM, an additional singlet superfield $\hat{Y}$ is
introduced, while the term $\mu(\hat{H}_1\hat{H}_2)$ is replaced
by $\lambda\hat{Y}(\hat{H}_1\hat{H}_2)+(\varkappa/3)\hat{Y}^3$. A
spontaneous breakdown of gauge symmetry leads to the emergence of
the vacuum expectation value $\langle Y\rangle=y/\sqrt{2}$ of the
field $Y$ and to generation an effective $\mu$--term
($\mu=\lambda\langle Y\rangle$). The resulting superpotential of
the nonminimal SUSY model is invariant under discrete
transformations of the $Z_3$ group \cite{8}. The $Z_3$ symmetry of
the superpotential of the observable sector naturally arises in
string models, where all observable fields are massless in the
limit of exact supersymmetry.

Upon the introduction of the neutral field $Y$ in the
superpotential of the NMSSM, there arises the corresponding
$F$--term in the potential of interaction of Higgs fields. As a
result, an upper limit on the mass of the lightest Higgs boson
becomes higher than that in the MSSM:
\begin{equation}
m_h\le\sqrt{\frac{\lambda^2}{2}v^2\sin^2 2\beta+ M_Z^2\cos^2
2\beta+\Delta}\, . \label{5}
\end{equation}
In the tree approximation ($\Delta=0$), relation (\ref{5}) was
obtained in \cite{9}. For $\lambda\to 0$, the expressions for the
upper limit in the MSSM and in the NMSSM coincide, after the
substitution of $\lambda y/\sqrt{2}$ for $\mu$. The Higgs sector
of the nonminimal SUSY model and one--loop corrections to it were
studied in \cite{10},\cite{11}. In \cite{12}, the upper limit on
the mass of the lightest Higgs boson within the NMSSM was
contrasted against the analogous limits in the minimal standard
and the minimal SUSY model.

From relation (\ref{5}), it follows that the upper limit on $m_h$
grows with increasing $\lambda(t_0)$. It should be noted that only
in the region of small values of $\tan\beta$ does it differ
significantly from the analogous limit in the MSSM. As to the
small $\tan\beta$ scenario, it is realised in the case of
sufficiently large values of $h_t(t_0)$. The growth of the Yukawa
coupling constants at the electroweak scale is accompanied by an
increase in $h_t(0)$ and $\lambda(0)$ at the Grand Unification
scale; therefore, the upper limit on the mass of the lightest
Higgs boson in the nonminimal SUSY model attains a maximum value
in the limit of strong Yukawa coupling, in which case both
$h_t^2(0)$ and $\lambda^2(0)$ are much greater than $g_i^2(0)$.

Unfortunately, we were unable to obtain a self--consistent
solution in the regime of strong Yukawa coupling within the NMSSM
featuring the minimal set of fundamental parameters. Moreover,
$Z_3$ symmetry, which makes it possible to avoid the problem of
the $\mu$--term in the nonminimal SUSY model, leads to the
emergence of three degenerate vacua in the theory. Immediately
following the phase transition at the electroweak scale, the
Universe is filled with three degenerate phases that must be
separated by domain walls. However, the hypothesis of a domain
structure of the vacuum is at odds with data from astrophysical
observations. An attempt at destroying $Z_3$ symmetry and the
domain structure of the vacuum by introducing nonrenormalizable
operators in the NMSSM Lagrangian leads to the appearance of
quadratic divergences -- that is, to the hierarchy problem
\cite{13}.

In order to avoid the domain structure of the vacuum and to obtain
a self--consistent solution in the regime of strong Yukawa
coupling, it is necessary to modify the nonminimal SUSY model. The
simplest way to modify the NMSSM is to introduce additional terms
in the superpotential of the Higgs sector,
$\mu(\hat{H}_1\hat{H}_2)$ and $\mu'\hat{Y}^2$ \cite{14}, that are
not forbidden by gauge symmetry. The additional bilinear terms in
the NMSSM superpotential destroy $Z_3$ symmetry, and domain walls
are not formed in such a theory. Upon the introduction of the
parameter $\mu$, it becomes possible to obtain the spectrum of
SUSY particles in the modified model; for a specific choice of
$\mu'$, the mass of the lightest Higgs boson reaches its upper
limit, taking the largest value at $\varkappa=0$. In analysing the
modified NMSSM, it is therefore reasonable to set the coupling
constant for the self--interaction of neutral superfields
$\hat{Y}$ to zero.

The MNSSM superpotential involves a large number of Yukawa
coupling constants. At $\tan\beta\sim 1$, they are all negligibly
small, however, with the exception of the $t$--quark Yukawa
coupling constant $h_t$ and the coupling constant $\lambda$, which
is responsible for the interaction of the superfield $\hat{Y}$
with the doublets $\hat{H}_1$ and $\hat{H}_2$. Thus, the total
superpotential of the modified NMSSM can be represented in the
form
\begin{equation}
W_{\text{MNSSM}}=\mu(\hat{H}_1\hat{H}_2)+\mu'\hat{Y}^2+
\lambda\hat{Y}(\hat{H}_1\hat{H}_2)+h_t(\hat{Q}\hat{H}_2)\hat{U}_R^c\,
. \label{6}
\end{equation}
Within SUGRA models, the terms in the superpotential (\ref{6})
that are bilinear in the superfields can be generated owing to the
term $(Z(H_1H_2)+Z'Y^2+h.c)$ in the K\"ahler potential
\cite{15},\cite{16} or owing to the nonrenormalized interaction of
fields from the observable and the hidden sector (this interaction
may be due to nonperturbative effects) \cite{16},\cite{17}.

Along with the parameters $\mu$ and $\mu'$, the masses of scalar
fields $m_i^2$, and the gaugino masses $M_i$ are also generated
upon a soft breakdown of supersymmetry. Moreover, a trilinear
coupling constant $A_i$ for the interaction of scalar fields is
associated in the total Lagrangian of the theory with each Yukawa
coupling constant, while a bilinear coupling constant $B$($B'$) is
associated there with the parameter $\mu$($\mu'$). The hypothesis
of universality of these constants at the scale $M_X$ makes it
possible to reduce their number to four: the scalar particle mass
$m_0$, the trilinear coupling constant $A$ and the bilinear
coupling constant $B_0$ for the interaction of scalar fields, and
the gaugino mass $M_{1/2}$.

\section{Analysis of the evolution of Yukawa couplings and determination of the quasi--fixed point}

The MNSSM parameters
\[ \lambda\,,\,\, h_t\,,\,\, \mu\,,\,\,
\mu'\,,\,\, m_0\,, \,\, A\,, \,\, B_0\, , \,\, M_{1/2} \]
specified at the Grand Unification scale evolve down to the
electroweak scale or the scale of supersymmetry breakdown. Their
renormalization is determined by the set of renormalization group
equations, these equations for the coupling constants $\lambda$,
$h_t$, $A_i$, $m_i^2$, and $M_i$ being coincident with the
corresponding renormalization group equations within the NMSSM
(see, for example, \cite{11}) if one sets $\varkappa=0$ in them.
The equations describing the evolution of $\mu$, $\mu'$, $B$, and
$B'$ within the modified NMSSM were obtained in \cite{14}.

Even in the one--loop approximation, the full system of
renormalization group equations is nonlinear, so that it is hardly
possible to solve it analytically. This set of equations can be
broken down into two subsets. The first subset includes equations
that describe the evolution of gauge and Yukawa coupling constants
and of parameters $\mu$ and $\mu'$. The second subset comprises
equations for the parameters of a soft breakdown of supersymmetry.

In studying the evolution of the Yukawa coupling constants, it is
convenient to introduce, instead of the constants $h_t$,
$\lambda$, and $g_i$, the ratios
\[
\rho_t(t)=\frac{Y_t(t)}{\tilde{\alpha}_3(t)}\, ,\quad
\rho_{\lambda}(t)=\frac{Y_{\lambda}(t)}{\tilde{\alpha}_3(t)}\,
,\quad \rho_1(t)=\frac{\tilde{\alpha}_1(t)}{\tilde{\alpha}_3(t)}\,
,\quad \rho_2(t)=\frac{\tilde{\alpha}_2(t)}{\tilde{\alpha}_3(t)}\,
,
\]
where $Y_{\lambda}(t)=\lambda^2(t)/(4\pi)^2$. The region of
admissible values of the Yukawa coupling constants at the
electroweak scale is bounded by the quasi--fixed (or Hill) line.
Beyond this region, solutions to the renormalization group
equations for $Y_i(t)$ develop a Landau pole below the scale
$M_X$, so that perturbation theory becomes inapplicable for
$q^2\sim M_X^2$. The results of our numerical calculations are
presented in Fig. 1, whence one can see that, in the regime of
strong Yukawa coupling, all solutions for $Y_i(t)$ are attracted
to the Hill line, which intersects the $\rho_t$ axis at the point
whose coordinates $(\rho_\lambda(t_0),\rho_t(t_0))=(0,0.087)$
correspond to the quasi--fixed point in the MSSM.

In analysing the results of the numerical calculations (see Fig.
1), attention is captured by a pronounced nonuniformity in the
distribution of solutions to the renormalization group equations
along the quasi--fixed line. The main reason behind this is that,
in the regime of strong Yukawa coupling, solutions are attracted
not only to the quasi--fixed but also to the infrared fixed (or
invariant) line. The latter connects two fixed points. One of them
is the stable infrared fixed point for the set of renormalization
group equations within the MNSSM
$(\rho_t=7/18,\rho_\lambda=0,\rho_1=0,\rho_2=0)$ \cite{18}. As the
invariant line approaches this point, $\rho_\lambda\sim
(\rho_t-7/18)^{25/14}$. The other fixed point
$(\rho_\lambda/\rho_t)=1$ corresponds to large values of the
Yukawa coupling constants, $Y_t,Y_\lambda\gg\tilde{\alpha}_i$, in
which case the gauge coupling constants can be disregarded
\cite{19}. In the limiting case of $\rho_\lambda,\rho_t\gg 1$, the
asymptotic behaviour of the curve being studied is given by
\begin{equation}
\rho_{\lambda}=\rho_t-\frac{8}{15}-\frac{2}{75}\rho_1. \label{7}
\end{equation}
The infrared fixed lines and their properties in the minimal
standard and the minimal supersymmetric model were studied in
detail elsewhere \cite{20}.

With increasing initial values of the Yukawa coupling constants
$Y_t(0)$ and $Y_\lambda(0)$ at the Grand Unification scale, the
region where solutions are concentrated at the electroweak scale
shrinks abruptly and all solutions to the renormalization group
equations within the MNSSM are focused near the point of
intersection of the invariant and the quasi--fixed line:
\begin{equation}
\rho^{\text{QFP}}_t(t_0)=0.803\, ,\qquad
\rho^{\text{QFP}}_{\lambda}(t_0)=0.224\,. \label{8}
\end{equation}
This point can be considered as the quasi--fixed point for the set
of renormalization group equations for the modified NMSSM
\cite{21}.

Among subsidiary constraints that are frequently imposed in
studying supersymmetric models, we would like to mention the
unification of the Yukawa coupling constants for the $b$--quark
and the $\tau$--lepton at the scale $M_X$; this usually occurs in
minimal schemes for unifying gauge interactions -- for example, in
those that are based on the $SU(5)$, the $E_6$, or the $SO(10)$
group. The unification of $h_b$ and $h_\tau$ within the MNSSM is
realised only in the case where the constants satisfy a specific
relation between $Y_t$ and $Y_\lambda$. Integrating the
renormalization group equations and substituting
$R_{b\tau}(t_0)=m_b(t_0)/m_\tau(t_0)=1.61$, which corresponds to
$m_\tau(t_0)=1.78\text{~GeV}$ and $m_b(t_0)=2.86\text{~GeV}$, we
obtain
\begin{equation}
\frac{Y_t(0)}{Y_t(t_0)}=\left[\frac{R_{b\tau}(0)}{R_{b\tau}(t_0)}
\right]^{\sfrac{21}{2}}
\left[\frac{\tilde{\alpha}_3(t_0)}{\tilde{\alpha}_3(0)}\right]^{\sfrac{68}{9}}
\left[\frac{\tilde{\alpha}_2(t_0)}{\tilde{\alpha}_2(0)}\right]^{\sfrac{9}{4}}
\left[\frac{\tilde{\alpha}_1(t_0)}{\tilde{\alpha}_1(0)}\right]^{\sfrac{463}{396}}
\left[\frac{Y_{\lambda}(0)}{Y_{\lambda}(t_0)}\right]^{\sfrac{1}{4}}\approx
3.67
\left[\frac{Y_{\lambda}(0)}{Y_{\lambda}(t_0)}\right]^{\sfrac{1}{4}}.
\label{9}
\end{equation}
The results obtained here indicate that $b-\tau$ unification is
possible under the condition that $Y_t(0)\gg Y_t(t_0)$, which is
realised only in the regime of strong Yukawa coupling. By varying
the running mass $m_b(m_b)$ of the $b$--quark from $4.1$ to
$4.4\text{~GeV}$, we found that the equality of the Yukawa
coupling constants at the Grand Unification scale can be achieved
only at $\tan\beta\le 2$.

The possibility of unifying the Yukawa coupling constants within
the NMSSM was comprehensively studied in \cite{21},\cite{22}. The
condition $Y_b(0)=Y_\tau(0)$ imposes stringent constraints on the
parameter space of the model being studied. Since $h_b$ and
$h_\tau$ are small in magnitude at $\tan\beta\sim 1$, they can be
generated, however, owing to nonrenormalizable operators upon a
spontaneous breakdown of symmetry at the Grand Unification scale.
In this case, $h_b$ and $h_\tau$ may be different. In studying the
spectrum of superpartners below, we will not therefore assume that
$R_{b\tau}(0)=1$.

\section{Renormalization of the soft SUSY breaking parameters}

If the evolution of gauge and Yukawa coupling constants is known,
the remaining subset of renormalization group equations within the
MNSSM can be treated as a set of linear differential equations for
the parameters of a soft breakdown of supersymmetry. For universal
boundary conditions, a general solution for the trilinear coupling
constants $A_i(t)$ and for the masses of scalar fields $m_i^2(t)$
has the form
\begin{equation}
\begin{gathered}
A_i(t)=e_i(t)A+f_i(t)M_{1/2}\, ,\\
\mathfrak{M}_i^2(t)=a_i(t)m_0^2+b_i(t)M_{1/2}^2+c_i(t)AM_{1/2}+d_i(t)A^2\,
.
\end{gathered}
\label{10}
\end{equation}
The functions $e_i(t)$, $f_i(t)$, $a_i(t)$, $b_i(t)$, $c_i(t)$,
and $d_i(t)$, which determine the evolution of $A_i(t)$ and
$m_i^2(t)$, remain unknown. The results of our numerical
calculations reveal that these functions greatly depend on the
choice of Yukawa coupling constants at the scale $M_X$.

In analysing the behaviour of solutions to the renormalization
group equations in the regime of strong Yukawa coupling, it is
more convenient to consider, instead of the squares of the scalar
particle masses, their linear combinations
\begin{equation}
\begin{split}
\mathfrak{M}_t^2(t)&=m_2^2(t)+m_Q^2(t)+m_U^2(t)\, ,\\
\mathfrak{M}_{\lambda}^2(t)&=m_1^2(t)+m_2^2(t)+m_y^2(t)\, .\\
\end{split}
\label{11}
\end{equation}
For the universal boundary conditions, solutions to the
differential equations for $\mathfrak{M}_i^2(t)$ can be
represented in the same form as the solutions for $m_i^2(t)$ (see
(\ref{10})); that is
\begin{equation}
\mathfrak{M}_i^2(t)=3\tilde{a}_i(t)m_0^2+\tilde{b}_i(t)M_{1/2}^2+
\tilde{c}_i(t)A M_{1/2}+\tilde{d}_i(t)A^2. \label{12}
\end{equation}
Since the homogeneous equations for $A_i(t)$ and
$\mathfrak{M}_i^2(t)$ have the same form, the functions
$\tilde{a}_i(t)$ and $e_i(t)$ coincide.

With increasing $Y_i(0)$, the functions $e_i(t_0)$, $c_i(t_0)$,
and $d_i(t_0)$ decrease and tend to zero in the limit
$Y_i(0)\to\infty$. Concurrently, $A_t(t)$, $A_\lambda(t)$,
$\mathfrak{M}^2_t(t)$, and $\mathfrak{M}^2_\lambda(t)$ become
independent of $A$ and $m_0^2$, while relations (\ref{10}) and
(\ref{12}) are significantly simplified. This behaviour of the
solutions in question implies that, as the solutions to the
renormalization group equations for the Yukawa coupling constants
approach quasi--fixed points, the corresponding solutions for
$A_i(t)$ and $\mathfrak{M}_i^2(t)$ also approach the quasi--fixed
points whose coordinates are \cite{23}
\begin{equation}
\begin{aligned} \rho_{A_t}^{\text{QFP}}(t_0)&\approx 1.77,
&\rho_{\mathfrak{M}^2_t}^{\text{QFP}}(t_0)&\approx 6.09,\\
\rho_{A_{\lambda}}^{\text{QFP}}(t_0)&\approx -0.42,\qquad
&\rho_{\mathfrak{M}^2_{\lambda}}^{\text{QFP}}(t_0)&\approx -2.28,
\end{aligned}
\label{13}
\end{equation}
where $\rho_{A_i}(t)=A_i(t)/M_{1/2}$ and
$\rho_{\mathfrak{M}_i^2}(t)=\mathfrak{M}_i^2/M_{1/2}^2$. At the
same time, the functions $a_i(t)$ approach some constants
independent of $t$ and $Y_i(0)$:
\begin{equation}
\begin{gathered} a_y(t)\to 1/7, \quad a_1(t)\to 4/7, \quad a_2(t)\to
-5/7,\\ a_u(t)\to 1/7, \quad a_q(t)\to 4/7. \end{gathered}
\label{14}
\end{equation}

In the case of nonuniversal boundary conditions at $Y_t(0)\simeq
Y_\lambda(0)$, the required solution to the renormalization group
equations for $A_i(t)$ and $\mathfrak{M}_i^2(t)$ can be
represented as \cite{23}
\begin{equation}
\begin{split}
\binom{A_t(t)}{A_\lambda(t)}&=\alpha_1\binom{v_{11}(t)}{v_{21}(t)}(\epsilon_t(t))^{\lambda_1}+
\alpha_2\binom{v_{12}(t)}{-3v_{22}(t)}(\epsilon_t(t))^{\lambda_2}+\dots,\\
\binom{\mathfrak{M}^2_t(t)}{\mathfrak{M}^2_\lambda(t)}&=\beta_1\binom{v_{11}(t)}{v_{21}(t)}(\epsilon_t(t))^{\lambda_1}+
\beta_2\binom{v_{12}(t)}{-3v_{22}(t)}(\epsilon_t(t))^{\lambda_2}+\dots,
\end{split}
\label{15}
\end{equation}
where $\alpha_i$ and $\beta_i$ are constants of integration that
can be expressed in terms of $A_t(0)$, $A_\lambda(0)$,
$\mathfrak{M}^2_t(0)$, and $\mathfrak{M}^2_\lambda(0)$;
$\epsilon_t(t)=Y_t(t)/Y_t(0)$; $\lambda_1=1$ and $\lambda_2=3/7$.
The functions $v_{ij}(t)$ are weakly dependent on the Yukawa
coupling constants at the scale $M_X$, and $v_{ij}(0)=1$. In
equations (\ref{15}), we have omitted terms proportional to
$M_{1/2}$, $M_{1/2}^2$, $A_i(0)M_{1/2}$, and $A_i(0)A_j(0)$.

With increasing $Y_t(0)\simeq Y_\lambda(0)$, the dependence of
$A_i(t_0)$ and $\mathfrak{M}_i^2(t_0)$ on $\alpha_1$ and $\beta_1$
quickly becomes weaker. The results of our numerical analysis that
are displayed in Fig. 2 indicate that, for
$h_t^2(0)=\lambda^2(0)=20$ and boundary conditions uniformly
distributed in the $(A_t,A_\lambda)$ and the
$(\mathfrak{M}_t^2,\mathfrak{M}_\lambda^2)$ plane, the solutions
to the renormalization group equations for the parameters of the
soft SUSY breaking in the vicinity of the quasi--fixed point are
concentrated near some straight lines. The equations of these
straight lines can be obtained by setting $A_\lambda(0)=-3A_t(0)$
and $\mathfrak{M}_\lambda^2(0)=-3\mathfrak{M}_t^2(0)$ (that is,
$\alpha_1=\beta_1=0$) at the Grand Unification scale. As a result,
we find that, at the electroweak scale, the parameters of a soft
breakdown of supersymmetry satisfy the relations
\begin{equation}
\begin{gathered}
A_t+0.137(0.147)A_{\lambda}=1.70M_{1/2},\\
\mathfrak{M}^2_t+0.137(0.147)\mathfrak{M}^2_{\lambda}=5.76
M_{1/2}^2 .
\end{gathered}
\label{16}
\end{equation}
The equation for $\mathfrak{M}_i^2$ has been obtained for all
$A_i(0)$ set to zero. In relations (\ref{16}), the coefficients
obtained by fitting the results of our numerical calculations (see
Fig. 2) are indicated parenthetically. As the Yukawa coupling
constants approach quasi--fixed points, the two sets of
coefficients in (\ref{16}) approach fast each other and, at
$Y_i(0)\sim 1$, become virtually coincident.

\section{Spectrum of SUSY particles and Higgs bosons}

Let us now proceed to study of the spectrum of the superpartners
of observable particles and Higgs bosons in the vicinity of the
quasi--fixed point within the MNSSM. The Yukawa coupling constants
$h_t$ and $\lambda$ are determined here by relations (\ref{8}).
The value of $\tan\beta$ can be calculated by formula (\ref{3}).
In the regime of the infrared quasi--fixed point at
$m_t(M_t^{\text{pole}})=165\text{~GeV}$ the result is
$\tan\beta=1.88$.

The remaining fundamental parameters of the MNSSM must be chosen
in such a way that a spontaneous breakdown of $SU(2)\otimes U(1)$
gauge symmetry occur at the electroweak scale. The position of the
physical minimum of the potential representing the interaction of
Higgs fields is determined by solving the set of nonlinear
algebraic equations
\begin{equation}
\frac{\partial V(v_1,v_2,y)}{\partial v_1}=0,\quad \frac{\partial
V(v_1,v_2,y)}{\partial v_2}=0,\quad \frac{\partial
V(v_1,v_2,y)}{\partial y}=0, \label{17}
\end{equation}
where $V(v_1,v_2,y)$ is the effective potential of interaction of
Higgs fields within the MNSSM \cite{14}.

Since the vacuum expectation value $v$ and $\tan\beta$ are known,
the set of equations (\ref{17}) can be used to determine the
parameters $\mu$ and $B_0$ and to compute the vacuum expectation
value $\langle Y\rangle$. Instead of $\mu$, it is convenient to
introduce here $\mu_{\text{eff}}=\mu+\lambda y/\sqrt{2}$. The sign
of $\mu_{\text{eff}}$ is not fixed in solving the set of equations
(\ref{17}); it must be considered as a free parameter of the
theory. The results obtained in this way for the vacuum
expectation value $y$, the parameters $\mu_{\text{eff}}$ and
$B_0$, and the spectrum of particles within the modified NMSSM
depend on the choice of $A$, $m_0$, $M_{1/2}$, and $\mu'$.

It is of particular interest to analyse the spectrum of particles
in that region of the parameter space of the MNSSM where the mass
of the lightest Higgs boson is close to its theoretical upper
limit, since the remaining part of the parameter space is
virtually ruled out by the existing experimental data. For each
individual set of the parameters $A$, $m_0$, and $M_{1/2}$, the
mass of the lightest Higgs boson reaches the upper bound on itself
at a specific choice of $\mu'$. It is precisely at these values of
the parameter $\mu'$ that we have calculated the particle spectrum
presented in Tables 1 and 2. On the basis of our numerical results
given there, one can judge the effect of the fundamental constants
$A$, $m_0$, and $M_{1/2}$ on the spectrum of the superpartners of
the $t$--quark ($m_{\tilde{t}_{1,2}}$), of the gluino ($M_3$), of
the neutralino ($m_{\tilde{\chi}_{1,\dots,5}}$), of the chargino
($m_{\tilde{\chi}^\pm_{1,2}}$), and of the Higgs bosons ($m_h$,
$m_H$, $m_S$, $m_{A_{1,2}}$). For each set of the aforementioned
parameters, we quote the mass of the lightest Higgs boson
according to the calculations in the one-- and the two--loop
approximation, along with the corresponding values of
$\mu_{\text{eff}}$, $B_0$, $y$, and $\mu'$. As can be seen from
the data displayed in Tables 1 and 2, the qualitative pattern of
the spectrum within the MNSSM undergoes no changes in response to
variations of the parameters $A$ and $m_0$ within reasonable
limits.

The CP--even Higgs boson ($m_S$), which corresponds to the neutral
field $Y$ is the heaviest particle in the spectrum of the modified
NMSSM, while the neutralino ($m_{\chi_5}$) is the heaviest fermion
there, the main contribution to its wave function coming from the
superpartner of the field $Y$. With increasing $m_0^2$ the masses
of the squarks, the Higgs bosons, and the heavy chargino and
neutralino grow, whereas the spectrum of extremely light particles
remains unchanged. Since the dependence of the parameters of a
soft breakdown of supersymmetry on $A$ disappears at the
electroweak scale in the regime of strong Yukawa coupling, the
parameters $\mu_{\text{eff}}$, $B$, and $\mu'$, together with the
spectrum of the superpartners of observable particles and the mass
of the lightest Higgs boson, undergo only slight changes in
response to a variation of the trilinear coupling constant for the
interaction of scalar fields from $-M_{1/2}$ to $M_{1/2}$. Despite
this, the $A$ dependence of masses of one of the CP--even ($m_S$)
and two CP--odd ($m_{A_{1,2}}$) Higgs bosons survives. It is due
primarily to the fact that the bilinear coupling constant $B'$ for
the interaction of neutral scalar fields is proportional to $A$.
It should be noted in addition that, for a specific choice of
fundamental parameters (in particular, of the parameter $A$), the
mass of the lightest CP--odd Higgs boson may prove to be about
$100\text{~GeV}$ or less. However, this Higgs boson takes
virtually no part in electroweak interactions, since the main
contribution to its wave function comes from the CP--odd component
of the field $Y$. Therefore, attempts at experimentally detecting
it run into problems.

Loop corrections play an important role in calculating the mass of
the lightest Higgs boson. Their inclusion results in that the mass
of the lightest Higgs boson proves to be greater for
$\mu_{\text{eff}}<0$ than for $\mu_{\text{eff}}>0$. This is
because $m_h$ grows as the mixing in he sector of the
superpartners of the $t$--quark ($\tilde{t}_R$ and $\tilde{t}_L$)
becomes stronger. The point is that the mixing of $\tilde{t}_R$
and $\tilde{t}_L$ is determined by the quantity
$X_t=A_t+\mu_{\text{eff}}/\tan\beta$ and is therefore greater in
magnitude for $\mu_{\text{eff}}<0$ since $A_t<0$. It should also
be noted that the inclusion of two--loop corrections leads to a
reduction of $m_h$ by approximately $10\text{~GeV}$. The mass of
the lightest Higgs boson depends only slightly on $A$ and $m_0$,
because the squark masses depend slightly on the corresponding
fundamental parameters (see Tables 1 and 2). The value of $m_h$ is
determined primarily by the supersymmetry breaking scale $M_S$ --
that is, by the quantity $M_3$. From our numerical results quoted
in Tables 1 and 2, one can see that, at
$m_t(M_t^{\text{pole}})=165\text{~GeV}$ and $M_3\le 2\text{~TeV}$,
the mass of the lightest Higgs boson does not exceed
$127\text{~GeV}$.

\section{Conclusions}

We have studied coupling constant renormalization and the spectrum
of particles within the modified Next--to--Minimal Supersymmetric
Standard Model (MNSSM). We have shown that, in the regime of
strong Yukawa coupling, solutions to the renormalization group
equations for $Y_i(t)$ are attracted to the Hill line and that,
under specific conditions $b-\tau$ unification is realised at the
scale $M_X$. In the limit $Y_i(0)\to\infty$, all solutions for the
Yukawa coupling constants are concentrated in the vicinity of the
quasi--fixed point that is formed in the space of the Yukawa
coupling constants as the result of intersection of the invariant
and the Hill line.

As the Yukawa coupling constants approach the quasi--fixed point,
the corresponding trilinear coupling constants and combinations
(\ref{11}) of the scalar particle masses cease to depend on the
boundary conditions at the scale $M_X$. In the case of
nonuniversal boundary conditions, $A_i(t)$ and
$\mathfrak{M}_i^2(t)$ are attracted to straight lines in the space
spanned by the parameters of a soft breakdown of supersymmetry
and, with increasing $Y_i(0)$, approach the quasi--fixed points,
moving along these straight lines.

We have analysed the spectrum of particles in the infrared
quasi--fixed point regime of the MNSSM. The CP--even Higgs boson,
which corresponds to the neutral field $Y$, is the heaviest
particle in this spectrum. At reasonable values of the parameters
of the model being studied, the gluinos, the squarks, and the
heavy Higgs bosons are much heavier than the lightest Higgs boson
and than the lightest chargino and the lightest neutralino as
well. This is not so only for one of the CP--odd Higgs bosons
whose mass changes within a wide range in response to variations
in the fundamental parameters of the MNSSM. In the vicinity of the
quasi--fixed point at $m_t(M_t^{\text{pole}})=165\text{~GeV}$, the
mass of the lightest Higgs boson does not exceed $127\text{~GeV}$.

\section*{Acknowledgements}

The authors are grateful to M. I. Vysotsky, D. I. Kazakov, L. B.
Okun, and K. A. Ter--Martirosyan for stimulating questions and
comments, and P. M. Zerwas and D. J. Miller for fruitful
discussions. R. B. Nevzorov is indebted to DESY Theory Group for
hospitality extended to him where the main part of this paper was
made.

This work was supported by the Russian Foundation for Basic
Research (RFBR), projects \#\# 00-15-96786, 00-15-96562,
02-02-17379.

\newpage

{\bfseries Table 1.} Particle spectrum in the vicinity of the
quasi--fixed point in the MNSSM at
$m_t(M_t^{\text{pole}})=165\text{~GeV}$, $\tan\beta\approx 1.883$,
and $\mu_{\text{eff}}>0$ depending upon the choice of fundamental
parameters $A$, $m_0$, and $M_{1/2}$ (all parameters and masses
are given in GeV).

\vspace*{5mm}

\begin{center}

\begin{tabular}{|c|c|c|c|c|c|c|}
\hline $m^2_0$&0&$M^2_{1/2}$&0&0&0&0\\ \hline
$A$&0&0&$-M_{1/2}$&$0.5 M_{1/2}$&0&0\\ \hline
$M_{1/2}$&-392.8&-392.8&-392.8&-392.8&-785.5&-196.4\\ \hline
$m_t(t_0)$&165&165&165&165&165&165\\ \hline
$\tan\beta$&1.883&1.883&1.883&1.883&1.883&1.883\\ \hline
$\mu_{\text{eff}}$&728.6&841.7&726.8&730.1&1361.2&380.4\\ \hline
$B_0$&-1629.1&-1935.4&-1260.0&-1813.2&-3064.4&-861.8\\ \hline
$y$&-0.00037&-0.00021&-0.00043&-0.00035&-0.00006&-0.00233\\ \hline
$\mu'(t_0)$&-1899.8&-2176.7&-1905.9&-1898.3&-3544.6&-993.1\\
\hline $\mathbf{m_h (t_0)}$&{\bf 125.0}&{\bf 125.1}&{\bf
125.0}&{\bf 125.0}& {\bf 134.9}&{\bf 114.8}\\ {\bf
(1--loop)}&&&&&&\\ \hline $\mathbf{m_h (t_0)}$&{\bf 118.4}&{\bf
118.5}&{\bf 118.4}&{\bf 118.4}& {\bf 123.2}&{\bf 111.9}\\ {\bf
(2--loop)}&&&&&&\\ \hline $M_3(1\text{~TeV})$&
1000&1000&1000&1000&2000&500 \\ \hline
$m_{\tilde{t}_1}(1\text{~TeV})$&840.6&889.7&841.1&840.3&1652.0&447.4\\
\hline
$m_{\tilde{t}_2}(1\text{~TeV})$&695.1&713.6&696.6&694.3&1366.2&371.6\\
\hline $m_H(1\text{~TeV})$&898.5&1080.5&895.4&900.3&1691.0&468.8\\
\hline $m_S(1\text{~TeV})$&
2623.4&3034.3&2452.2&2706.0&4901.7&1378.0 \\ \hline
$m_{A_1}(1\text{~TeV})$&953.9&1113.8&1245.7&925.2&1722.6&538.2\\
\hline
$m_{A_2}(1\text{~TeV})$&704.3&762.7&872.0&318.2&1366.2&302.2\\
\hline
$m_{\tilde{\chi}_1}(t_0)$&164.6&164.4&164.6&164.6&326.9&84.3\\
\hline
$m_{\tilde{\chi}_2}(t_0)$&327.8&327.6&327.8&327.8&649.4&170.1\\
\hline
$m_{\tilde{\chi}_3}(1\text{~TeV})$&755.1&870.8&753.3&756.7&1404.2&400.9\\
\hline
$|m_{\tilde{\chi}_4}(1\text{~TeV})|$&755.9&872.6&755.1&758.4&1405.0&404.3\\
\hline
$|m_{\tilde{\chi}_5}(1\text{~TeV})|$&1931.8&2212.3&1938&1930.3&3599.0&1015.4\\
\hline
$m_{\tilde{\chi}^{\pm}_1}(t_0)$&327.8&327.6&327.8&327.8&649.4&169.9\\
\hline
$m_{\tilde{\chi}^{\pm}_2}(1\text{~TeV})$&757.0&872.6&755.2&758.5&1405.2&404.5\\
\hline
\end{tabular}

\end{center}

\newpage

{\bfseries Table 2.} Particle spectrum in the vicinity of the
quasi--fixed point in the MNSSM at
$m_t(M_t^{\text{pole}})=165\text{~GeV}$, $\tan\beta\approx 1.883$,
and $\mu_{\text{eff}}<0$ depending upon the choice of fundamental
parameters $A$, $m_0$, and $M_{1/2}$ (all parameters and masses
are given in GeV).

\vspace*{5mm}

\begin{center}

\begin{tabular}{|c|c|c|c|c|c|c|c|c|}
\hline $m^2_0$&0&$M^2_{1/2}$&0&0&0&0\\ \hline
$A$&0&0&$-M_{1/2}$&$M_{1/2}$&0&0\\ \hline
$M_{1/2}$&-392.8&-392.8&-392.8&-392.8&-785.5&-196.4\\ \hline
$m_t(t_0)$&165&165&165&165&165&165\\ \hline
$\tan\beta$&1.883&1.883&1.883&1.883&1.883&1.883\\ \hline
$\mu_{\text{eff}}$&-727.8&-840.9&-726.0&-731.2&-1360.7&-378.9\\
\hline $B_0$&1008&1320.3&1366.7&647.9&2050.4&495.8\\ \hline
$y$&-0.00149&-0.001&-0.00128&-0.00177&-0.00020&-0.0112\\ \hline
$\mu'(t_0)$&1671.5&1950.6&1656.8&1690.3&3172.7&857.8\\ \hline
${\bf m_h (t_0)}$&{\bf 134.1}&{\bf 134.9}&{\bf 134.0}& {\bf
134.2}&{\bf 143.1}&{\bf 124.1}\\ {\bf (1--loop)}&&&&&&\\ \hline
${\bf m_h (t_0)}$&{\bf 124.4}&{\bf 124.8}&{\bf 124.3}&{\bf 124.5}
&{\bf 127.2}&{\bf 119.6}\\ {\bf (2--loop)}&&&&&&\\ \hline
$M_3(1\text{~TeV})$&1000&1000&1000&1000&2000&500\\ \hline
$m_{\tilde{t}_1}(1\text{~TeV})$&890.2&935.6&890.5&889.8&1682.8&507.9\\
\hline
$m_{\tilde{t}_2}(1\text{~TeV})$&630.3&652.2&632.2&628.0&1328.1&283.5\\
\hline $m_H(1\text{~TeV})$&896.2&1078.5&893.5&899.3&1689.9&464.4\\
\hline
$m_S(1\text{~TeV})$&2147.4&2565.9&2309.2&1972.3&4126.5&1097.7\\
\hline
$m_{A_1}(1\text{~TeV})$&1123.2&1219.3&931.0&1437.9&1984.8&623.1\\
\hline
$m_{A_2}(1\text{~TeV})$&857.6&1017.8&545.0&886.9&1657.5&412.8\\
\hline
$m_{\tilde{\chi}_1}(t_0)$&160.0&160.5&160.0&160.0&324.4&74.9\\
\hline
$m_{\tilde{\chi}_2}(t_0)$&311.1&313.7&311.0&311.2&639.9&141.4\\
\hline
$|m_{\tilde{\chi}_3}(1\text{~TeV})|$&753.7&896.6&751.9&757.2&1403.4&398.5\\
\hline
$m_{\tilde{\chi}_4}(1\text{~TeV})$&764.7&878.1&763.0&768.1&1410.0&416.7\\
\hline
$m_{\tilde{\chi}_5}(1\text{~TeV})$&1700.7&1983.2&1685.8&1719.6&3221.8&879.1\\
\hline
$m_{\tilde{\chi}^{\pm}_1}(t_0)$&310.7&313.4&310.7&310.8&639.8&139.4\\
\hline
$m_{\tilde{\chi}^{\pm}_2}(1\text{~TeV})$&763.3&877.0&761.6&766.7&1409.1&414.5\\
\hline
\end{tabular}

\end{center}

\newpage

\section*{Figure captions}

{\bfseries Fig.~1.} Boundary conditions for the renormalization
group equations of the MNSSM at the scale $q=M_X$ uniformly
distributed in the $(\rho_t,\rho_\lambda)$ plane in a square $2\le
h_t^2(0),\lambda^2(0)\le 10$ -- Fig.1a, and the corresponding
values of the Yukawa couplings at the electroweak scale -- Fig.1b.
The thick and thin curves in Fig.1b represent, respectively, the
invariant and the Hill line. The dashed straight line in Fig.1b is
a fit of the values $(\rho_t(t_0),\rho_\lambda(t_0))$ for $20\le
h_t^2(0),\lambda^2(0)\le 100$.\\

{\bfseries Fig.~2.} Boundary conditions for the renormalization
group equations of the MNSSM at the Grand Unification scale
($t=0$) at $h_t^2(0)=\lambda^2(0)=20$ uniformly distributed in the
$(A_t/M_{1/2},A_\lambda/M_{1/2})$ plane -- Fig. 2a, and the
corresponding values of the trilinear couplings at the electroweak
scale ($t=t_0$) -- Fig. 2b. The straight line in Fig. 2b is a fit
of the values $(A_t(t_0),A_\lambda(t_0))$.\\

{\bfseries Fig.~3.} Boundary conditions for the renormalization
group equations of the MNSSM at the Grand Unification scale
($t=0$) at $h_t^2(0)=\lambda^2(0)=20$ and $A_t(0)=A_\lambda(0)=0$
uniformly distributed in the
$(\mathfrak{M}_t^2/M_{1/2}^2,\mathfrak{M}_\lambda^2/M_{1/2}^2)$
plane -- Fig. 3a, and the corresponding values of the trilinear
couplings at the electroweak scale ($t=t_0$) -- Fig. 3b. The
straight line in Fig. 3b is a fit of the values
$(\mathfrak{M}_t^2(t_0),\mathfrak{M}_\lambda^2(t_0))$.\\

\begin{landscape}
\noindent%
\includegraphics[height=93mm,keepaspectratio=true]{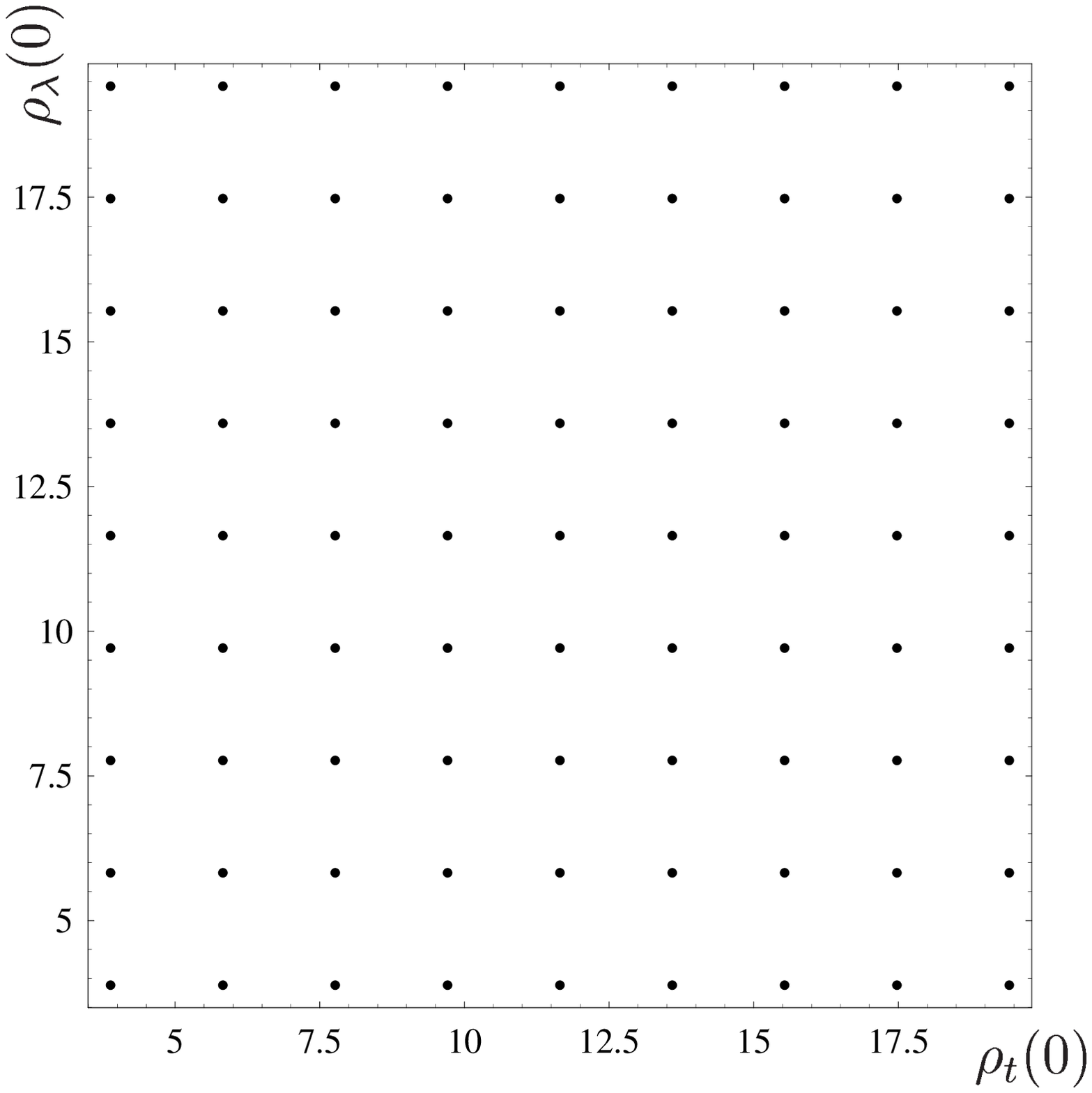}
\hfill\raisebox{-2mm}%
{\includegraphics[height=99mm,keepaspectratio=true]{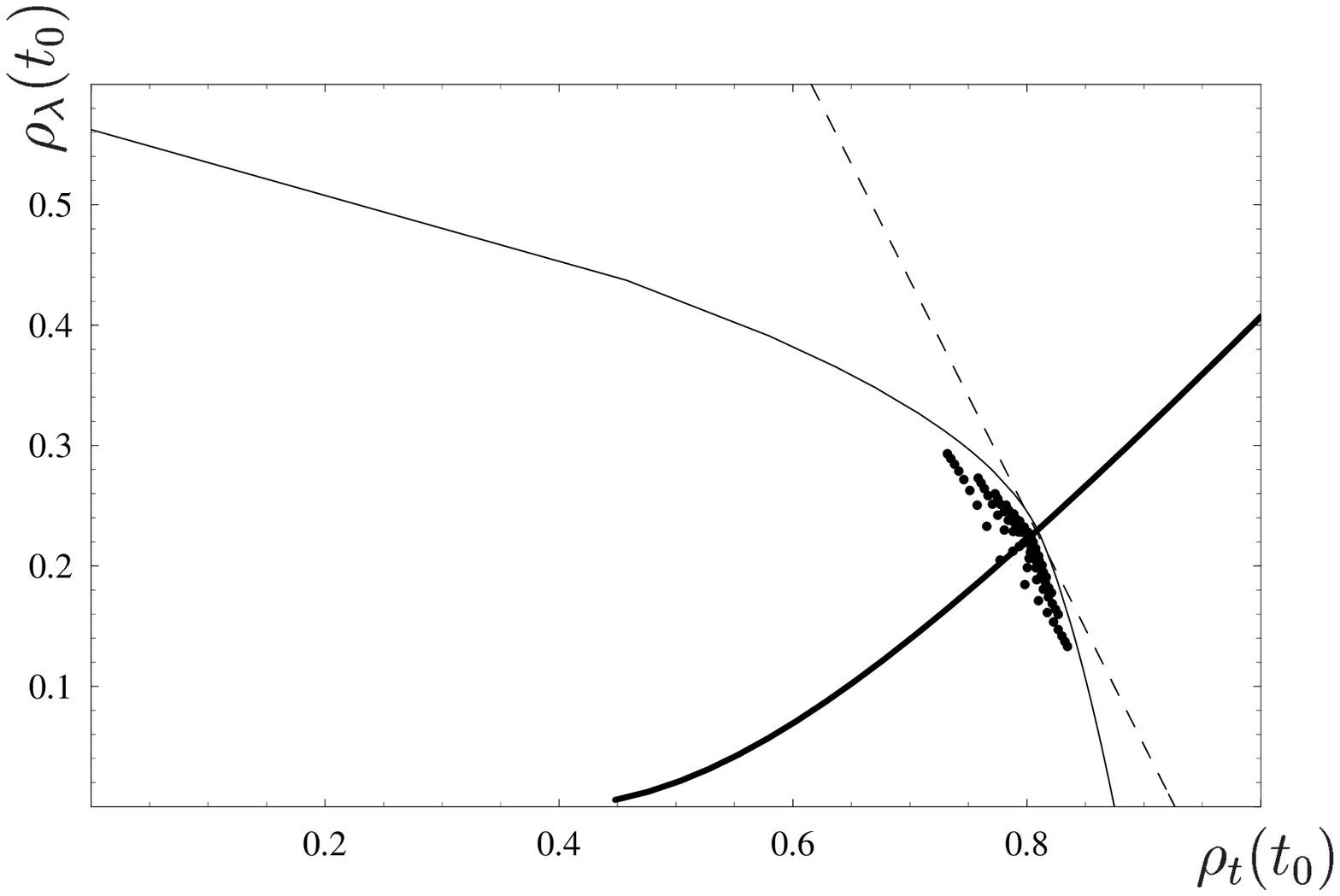}}

\vspace{5mm}\hspace*{32mm}{\large\bfseries
Fig.1a.}\hspace{107mm}{\large\bfseries Fig.1b.}

\end{landscape}

\begin{landscape}

\noindent\raisebox{3mm}{\includegraphics[height=92mm,keepaspectratio=true]{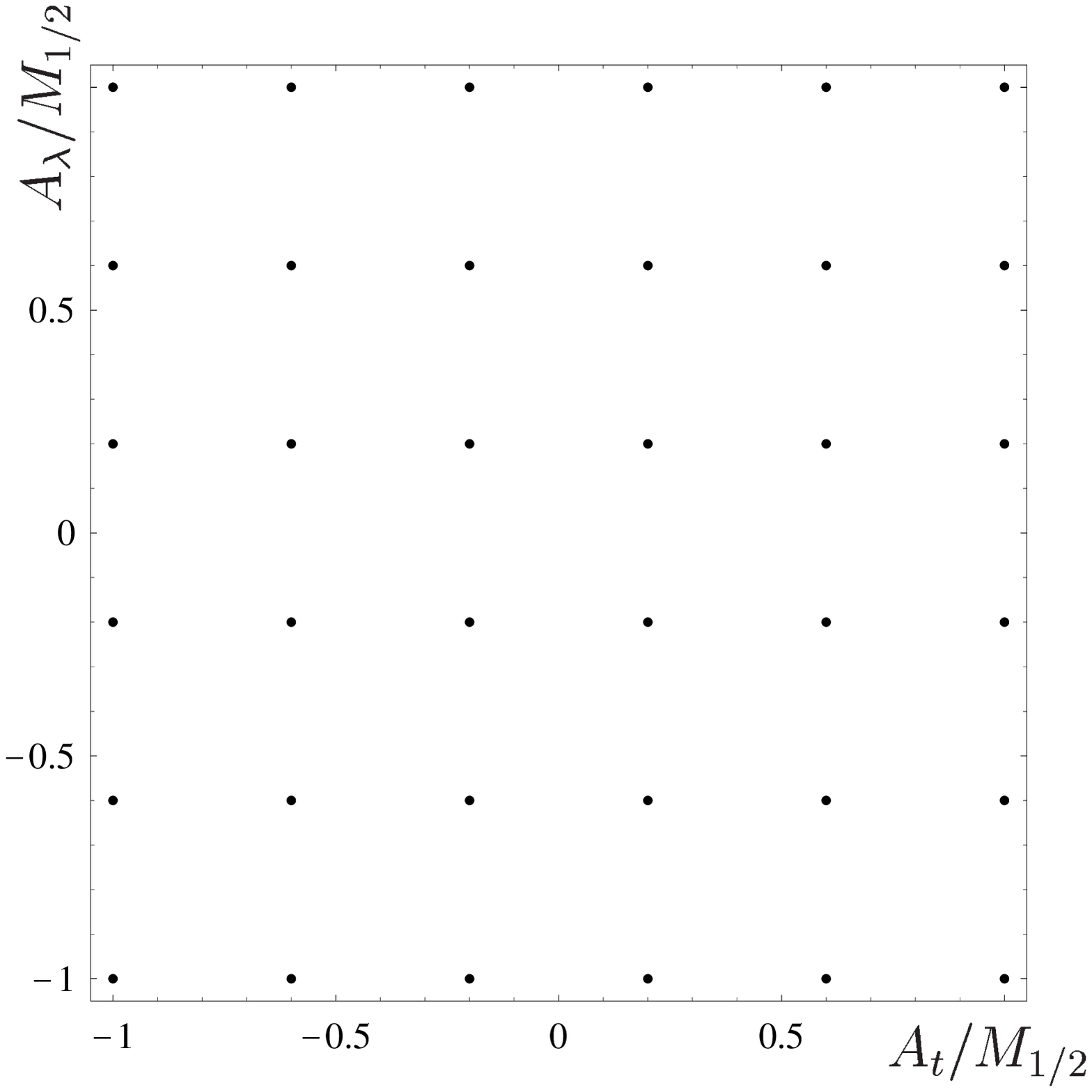}}
\hfill
\includegraphics[height=97mm,keepaspectratio=true]{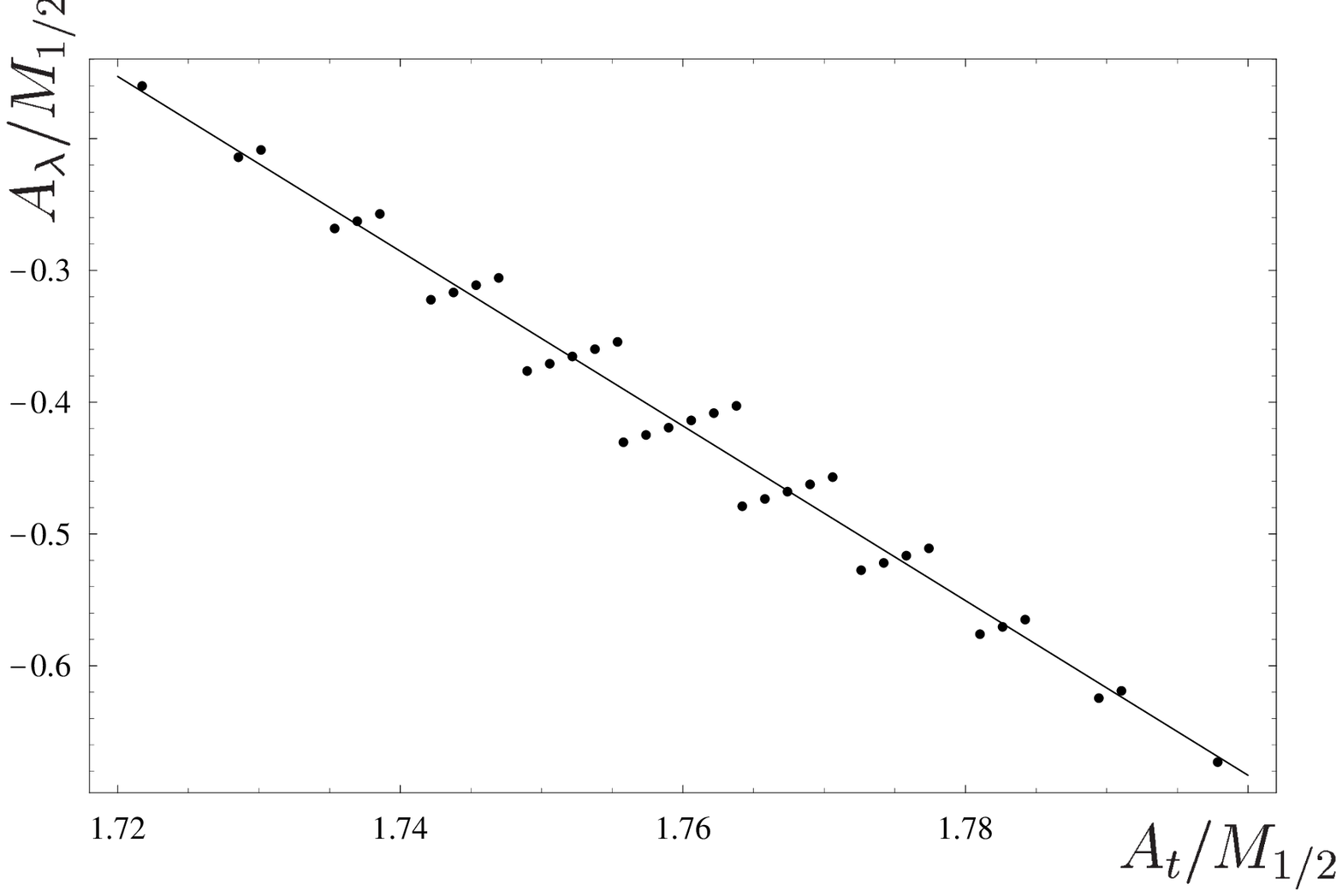}

\hspace*{32mm}{\large\bfseries
Fig.2a}\hspace*{109.5mm}{\large\bfseries Fig.2b.}

\end{landscape}

\begin{landscape}

\noindent\raisebox{2.5mm}{\includegraphics[height=92mm,keepaspectratio=true]{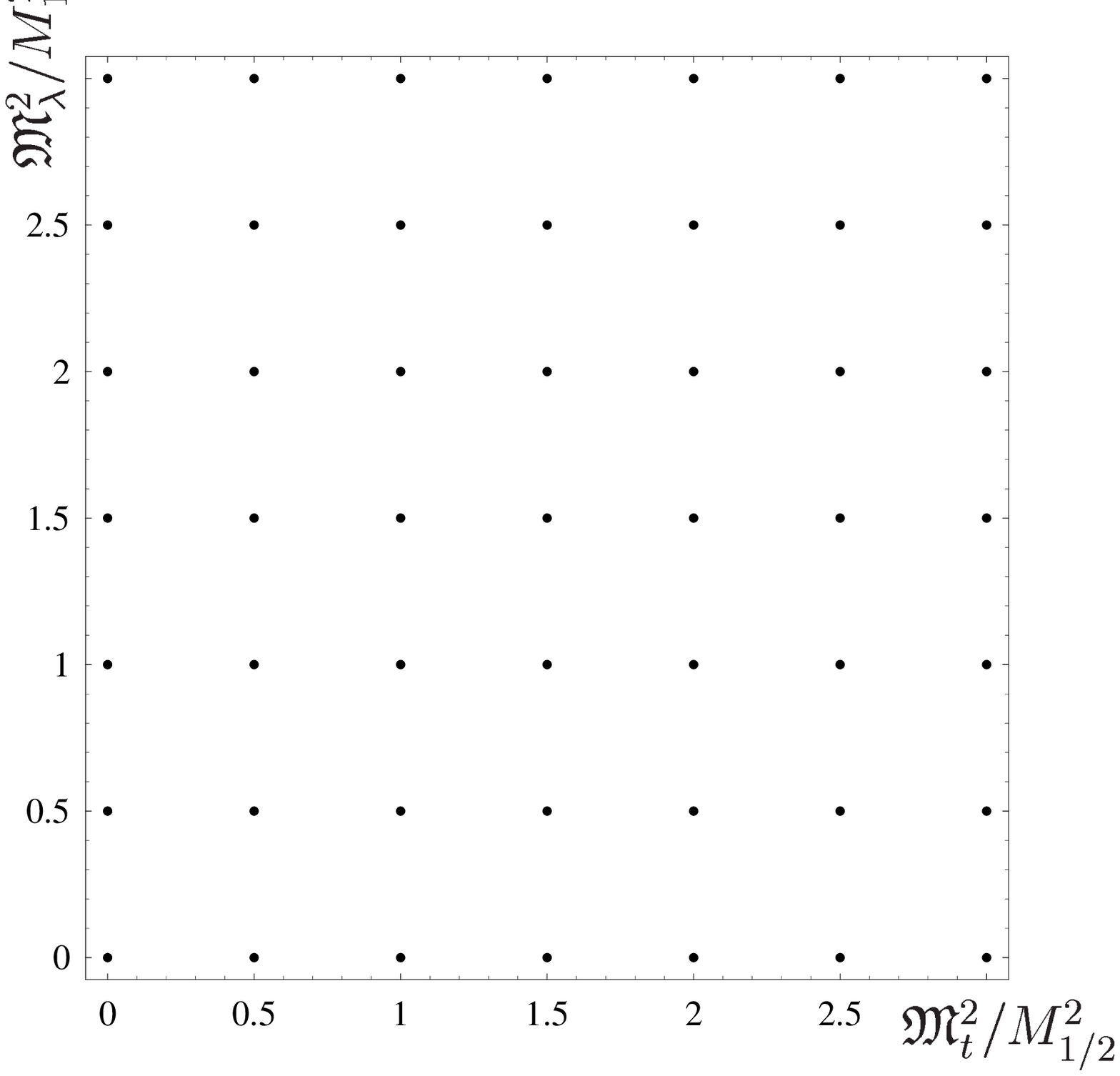}}
\hfill
\includegraphics[height=100mm,keepaspectratio=true]{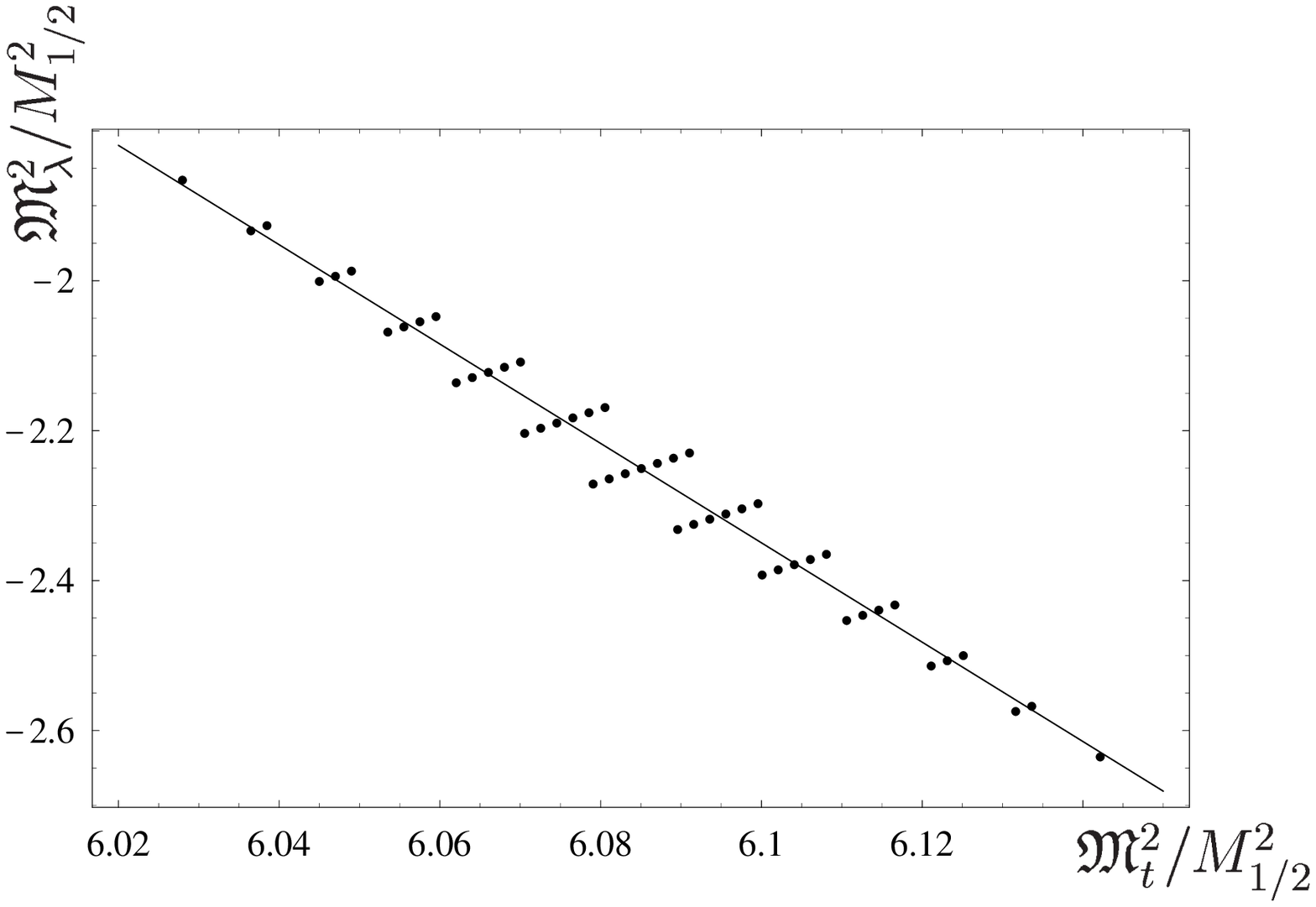}

\hspace*{30mm}{\large\bfseries
Fig.3a.}\hspace*{108.5mm}{\large\bfseries Fig.3b.}

\end{landscape}

\end{document}